\documentclass[prb,twocolumn,superscriptaddress,amsmath,amssymb,showpacs,floatfix,preprintnumbers]{revtex4}
\usepackage{amsmath}
\usepackage{graphicx}
\usepackage{dcolumn}
\usepackage{color}
\DeclareGraphicsExtensions{.png .jpg .pdf}

\begin{document}

\title{Boundary conditions for phosphorene nanoribbons in the continuum approach}

\author{D. J. P. de Sousa}\email{duarte.j@fisica.ufc.br}
\author{L. V. de Castro}\email{vieiradecastroluan@fisica.ufc.br}
\author{D. R. da Costa}\email{diego_rabelo@fisica.ufc.br}
\author{J. Milton Pereira Jr.}\email{pereira@fisica.ufc.br}
\affiliation{Departamento de F\'isica, Universidade Federal do Cear\'a, Caixa Postal 6030, Campus do Pici, 60455-900 Fortaleza, Cear\'a, Brazil}

\date{Published: 14 December 2016 - Phys. Rev. B \textbf{94}, 235415 (2016)}

\begin{abstract}
We investigate the energy spectrum of single layer black phosphorene nanoribbons (BPN) by means of a low-energy expansion of a recently proposed tight-binding model that describes electron and hole bands close to the Fermi energy level. Using the continuum approach, we propose boundary conditions based on sublattice symmetries for BPN with zigzag and armchair edges and show that our results for the energy spectra exhibit good agreement with those obtained by using the five-parameter tight-binding model. We also explore the behaviour of the energy gap versus the nanoribbon width $W$. Our findings demonstrate that band gap of armchair BPNs scale as $1/W^2$, while zigzag BPNs exhibit a $1/W$ tendency. We analyse the different possible combinations of the zigzag edges that result two-fold degenerate and non-degenerate edge states. Furthermore, we obtain expressions for the wave functions and discuss the limit of validity of such analytical model. 
\end{abstract}


\maketitle

\section{Introduction}

The great interest generated by the production of graphene, in 2004, has led to a search for graphene analogues that share its advantages, such as high electronic mobility, but not its shortcomings, such as the lack of an electronic band gap.\cite{CastroNetoReview,Misha1} Many of the new two-dimensional (2D) crystals investigated are obtained from layered materials which, similarly to graphene, can be mechanically exfoliated to obtain samples with few or single layers. \cite{Geim} That has resulted in the discovery of single layer crystals of Silicon (silicene) \cite{silicene}, Germanium (germanene) \cite{germanene}, as well as of a class of materials known as transition metal dichalcogenides \cite{Kis}. One of the most promising of this new crop of 2D materials is single layer Black Phosphorus (BP) \cite{bp1,bp2,bp3,bp4,bp5,bp6,bp7} which is a narrow gap semiconductor, also known as phosphorene, that has been found to display a carrier mobility in the range of $\approx 1000$ cm$^2$ V$^{-1}$s$^{-1}$, which is comparable to graphene, but with a highly anisotropic band structure. Moreover, few-layer phosphorene has been shown to display a tunable band gap\cite{bp2, bp5, Tran, Gomez, Dolui, Das, Kim, Katsnelson}, which is relevant for device applications. There is already a growing literature dealing with basic properties of phosphorene, as well as studies of possible technological applications.\cite{bp1, bp2, Yuan, Fazzio} A series of recent studies have obtained the electronic dispersion using approaches such as first principles calculations\cite{Rudenko, Dolui, Tran, Tran1, Carvalho}, tight-binding model \cite{Rudenko, Rudenko1}, ${\mathbf k} \cdot {\mathbf p}$ methods\cite{Zhou, Li}, and a long-wavelength approximation \cite{Milton}.

Following the example of graphene nanoribbons\cite{Brey, Enoki, Enoki1}, one can expect that the electronic spectrum and the transport properties of narrow phosphorene ribbons can be significantly distinct from the case of an infinite sample. Recent studies of BPNs have been based on a tight-binding approach\cite{Ezawa, Sisakht, Marko} and via first-principles simulations\cite{Tran1, Carvalho, Peng} that, while giving reasonably precise results for small structures, can become computationally expensive for larger structures. In addition, pure computational approaches are not appropriate to give physical insights into the basic mechanisms behind the results, which is of fundamental importance for a pure theoretical understanding. Therefore, in this work we investigate the electronic dispersion of phosphorene nanoribbons within the context of a continuum model based on the long-wavelength BP Hamiltonian \cite{Milton}. We obtain the boundary conditions that describe the behaviour of the envelope functions of the system and consider the effect of different edges (i.e. zigzag and armchair) on the spectrum. The band structures obtained analytically are compared with those ones calculated by using a five-hopping tight-binding Hamiltonian.  

The paper is organized as follows. In Sec. \ref{Sec.Model}, we describe the two theoretical models used to describe the charge carriers in single layer BPN: continuum approximation and the tight-binding approach. We propose the appropriate boundary conditions for armchair and zigzag BPN in Secs. \ref{Sec.AC} and \ref{Sec.ZZ}, respectively, and present the electronic band structures from BPNs associated to those boundary conditions. The electronic properties are found using both models and are compared in order to find the limit of accuracy of the analytical model. In Sec. \ref{Sec.GAP}, we analytically demonstrate for the scaling laws obeyed by the energy band gap for BPNs. Concluding remarks are reported in Sec. \ref{Sec.Conclusion}.

\section{Theoretical Model}\label{Sec.Model}

\begin{figure}[t]
\centerline{\includegraphics[width = \linewidth]{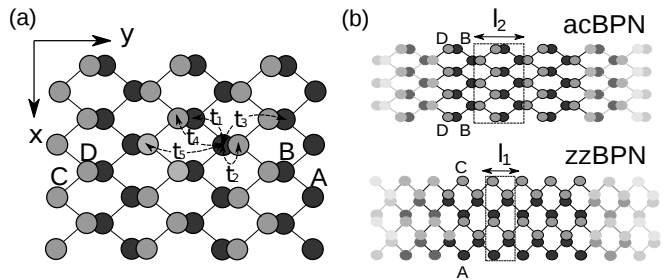}}
\caption{(Color online) (a) Top view of a phosphorene sheet, emphasizing the orientation of the system and the four atoms of the base. The sublattices $A$ and $B$ ($C$ and $D$) are at the bottom (top) of the monolayer structure, being represented by black (gray) circular symbols, respectively. $t_{1,2,3,4,5}$ are the five hopping energies for tight-binding model. (b) Armchair (acBPN) (top) and zigzag (zzBPN) (bottom) black phosphorene nanoribbons. The lattice parameters for zzBPN and acBPN are $l_{1} = 3.31$ \AA\ and $l_{2} = 4.19$ \AA, respectively.} 
\label{Fig1}
\end{figure}

In this section, we present the theoretical tools used to obtain the energy spectra of BPNs discussed in the following sections. Based on the reduced two-band model recently reported in Ref.~[\onlinecite{Milton}], the long-wavelength Hamiltonian for describing low-energy carriers in a phosphorene sheet around $\Gamma$ point reads in momentum space as
\begin{equation}\mathcal{H}' = \left(
\begin{array}{cc}
u_0 + \eta_x k_ {x}^{2} + \eta_y k_{y}^{2} & \delta + \gamma_x k_ {x}^{2} + \gamma_y k_{y}^{2} + i\chi k_{y} \\
\delta + \gamma_x k_ {x}^{2} + \gamma_y k_{y}^{2} - i\chi k_{y} & u_0 + \eta_x k_ {x}^{2} + \eta_y k_{y}^{2} 
\end{array}\right),
\label{PrincipalHamiltonian}
\end{equation}
which acts on the two component spinors $\Psi ' = [\phi_1 \ \  \phi_2]^{T}$, where $\phi_1 = \phi_A + \phi_D$ and $\phi_2 = \phi_B + \phi_C$, and the functions $\phi_{A,B,C,D}$ are the probability amplitudes for finding electrons on the atomic sites $A$, $B$, $C$ and $D$, respectively, which are related to the four phosphorus atoms that are contained in the unit cell of monolayer BP.\cite{Rudenko} The unitary transformation
\begin{equation}
U = \frac{1}{\sqrt 2}\left(
\begin{array}{cc}
1 & 1 \\
1 & -1
\end{array}\right),
\label{UnitaryTransformation}
\end{equation}
transforms the Hamiltonian (1) into a simpler form, which is given by
\begin{equation} U^{\dagger} \mathcal{H}' U = \mathcal{H} = \left(
\begin{array}{cc}
\alpha + \beta k_ {x}^{2} + \gamma k_{y}^{2} & i\chi k_{y} \\
-i\chi k_{y} & \bar{\alpha} + \bar{\beta}k_ {x}^{2} + \bar{\gamma} k_{y}^{2} 
\end{array}\right),
\label{eq1}
\end{equation}
with eigenstates
\begin{equation} 
U \Psi ' = \Psi = \left(\begin{array}{c}
\phi_{+} \\
\phi_{-} 
\end{array}\right) = \frac{1}{\sqrt{2}}
\left(\begin{array}{c}
\phi_{A} + \phi_{D} + \phi_{C} + \phi_{B} \\
\phi_{A} + \phi_{D} - \phi_{C} - \phi_{B} 
\end{array}\right),
\label{eq2}
\end{equation}
where the $x$ and $y$ coordinates correspond to the zigzag and armchair directions, respectively, as illustrated in Fig.~\ref{Fig1}(a). It must be emphasized that this is a long-wavelength approximation derived from a tight-binding model which fits \textit{ab-initio} calculations (see Ref.~[\onlinecite{Rudenko}]) which differs from that obtained by other authors that have used a $\textbf{k}\cdot \textbf{p}$ method\cite{Zhou, Li}. The tight-binding description gives a direct dependence of the eigenstate components with the sublattice amplitudes, which allows us to write the eigenstates for the two-band Hamiltonian, Eq.~(\ref{eq1}), in the form of Eq.~(\ref{eq2}), as shown in Ref.~[\onlinecite{Milton}]. Here $\alpha$ ($\bar{\alpha}$) = $u_{0} + \delta$ ($u_{0} - \delta$), $\beta$ ($\bar{\beta}$) = $\eta_{x} + \gamma_{x}$ ($\eta_{x} - \gamma_{x}$), $\gamma$ ($\bar{\gamma}$) = $\eta_{y} + \gamma_{y}$ ($\eta_{y} - \gamma_{y}$), with $u_{0} = -0.42$ eV, $\eta_{x} = 0.58$ eV$\cdot$\AA $^{2}$, $\eta_{y} = 1.01$ eV$\cdot$\AA $^{2}$, $\delta = 0.76$ eV, $\chi = 5.25$ eV$\cdot$\AA, $\gamma_{x} = 3.93$ eV$\cdot$\AA $^{2}$ and $\gamma_{y} = 3.79$ eV$\cdot$\AA $^{2}$. These parameters are the same ones used in Ref.~[\onlinecite{Milton}] and they include the contribution from the five hopping energies of the four-band tight-binding model for BP sheet and its lattice geometry. In this sense, those parameters incorporate a direct link between the microscopic tight-binding description and the continuum approximation. Note also that $E_c = \alpha = 0.34$ eV ($E_v = \bar{\alpha} = -1.18$ eV) is the conduction (valence) band edge, i.e. the conduction band minimum (the valence band maximum), leading to a band gap energy of $E_g = 1.52$ eV. Such energy gap is consistent with the recent  Photoluminescence measurements\cite{bp2}, and first principles simulations\cite{Rudenko, Tran, Tran1}.

Notice that the above two-band Hamiltonian (Eq.~(\ref{eq1})) and, consequently, the theoretical description considered in this paper, are obtained by taking advantage of the $D_{2h}$ point group invariance of the BP lattice\cite{Ezawa} which allows us to reduce the four-band model to the two-band model.   

The eigenvalue equation $H\Psi = E\Psi$ leads to two coupled second-order differential equations, given by
\begin{subequations}
\begin{align}
(\alpha - \beta \partial_{x}^{2} - \gamma \partial_{y}^{2})\phi_{+} + \chi\partial_ {y}\phi_{-} &= E\phi_{+},\label{eq3}\\
(\bar{\alpha} - \bar{\beta} \partial_{x}^{2} - \bar{\gamma} \partial_{y}^{2})\phi_{-} - \chi\partial_ {y}\phi_{+} &= E\phi_{-}.\label{eq4}
\end{align}
\end{subequations}
In the following sections, we solve the above differential equations (\ref{eq3}) and (\ref{eq4}) to study two different nanoribbon configurations, each corresponding to a specific type of edges: armchair (Sec.~\ref{Sec.AC}) and zigzag (Sec.~\ref{Sec.ZZ}).

In order to check the validity of the analytical results obtained via continuum model, we calculate the energy spectrum within the tight-binding description by using a software package for quantum transport called Kwant\cite{Kwant}. Such tight-binding Hamiltonian considered here has been recently proposed (see for more details Ref. [\onlinecite{Rudenko}]) and successfully used \cite{Ezawa, Marko, Zhang, Zhou} in some previous works in the literature. In this approach, the Hamiltonian for describing electrons in BPN can be written as
\begin{equation}\label{TBHamiltonian}
H_{TB} = \sum_{i\neq j}t_{ij} c_i^{\dagger}c_j,
\end{equation}
where $t_{ij}$ are the hopping energy terms between the $i$th and $j$th sites and the $c_{i}$ ($c_{i}^{\dagger}$) operators annihilate (create) an electron at site $i$. The summation runs over all the lattice sites of the BPN. The five most significant hopping integrals are illustrated in Fig.~\ref{Fig1} and suggested in Ref.~[\onlinecite{Rudenko}] as follows: $t_1 = -1.220$ eV, $t_2 = 3.665$ eV, $t_3 = -0.205$ eV, $t_4 = -0.105$ eV, and $t_5 = -0.055$ eV. The two distances $l_1$ and $l_2$ also correspond to the length of the unit cell of zzBPN and acBPN, respectively. It is important to point out that the adopted model presupposes first ($t_1$ and $t_2$), second ($t_4$) and third ($t_3$ and $t_5$) nearest-neighbor couplings and thereby each phosphorus atom is covalently coupled to three other ones, resulting in a puckered structure with a top view that resembles the honeycomb lattice of graphene.\cite{Milton} 

\section{Armchair phosphorene nanoribbons}\label{Sec.AC}

The geometry of an armchair phosphorene nanoribbon (acBPN) is illustrated at the top part of Fig.~\ref{Fig1}(b), where $l_{2} = 4.19$ \AA\ is the lattice parameter of the chosen unit cell along the $y$-axis, whereas the lattice is limited along the $x$-direction characterized by the width $W$. The phosphorus atoms in acBPNs are arranged in such a way that the type of termination sublattices are: (i) $A$ and $C$ in one edge and sublattices $B$ and $D$ in the other edge, or (ii) sublattices $A$ and $C$ ($B$ and $D$) present in both edges. In any case, by taking into account the symetries between sublattices $A/D$ and $B/C$ as a consequence of the $D_{2h}$ group invariance of the BP lattice, a pair of atoms of non-equivalent sublattices are always missing in both edges. In order to describe an armchair nanoribbon, we assume that the system has translational invariance only along the $y$-direction (see Fig.~\ref{Fig1}(a)). Thus, one can write $\phi \rightarrow \phi e^{ik_{y}y}$, and Eqs.~(\ref{eq3}) and (\ref{eq4}) are transformed to
\begin{subequations}
\begin{align}
(\alpha - \beta \partial_{x}^{2} + \gamma k_{y}^{2})\phi_{+} + i\chi k_ {y}\phi_{-} &= E\phi_{+}, \label{eq5}\\
(\bar{\alpha} - \bar{\beta} \partial_{x}^{2} + \bar{\gamma} k_{y}^{2})\phi_{-} - i \chi k_ {y}\phi_{+} &= E\phi_{-}.\label{eq6}
\end{align}
\end{subequations}
Decoupling the system of differential equations, we arrive at the fourth-order differential equation for the component $\phi_{+}$
\begin{equation}
(a\partial_{x}^{4} + b\partial_{x}^{2} + c)\phi_{+} = 0,
\label{eq7}
\end{equation}
and $\phi_{-}$ can be obtained from the relation
\begin{equation}
\phi_{-} = -\frac{i}{\chi k_{y}}(\beta \partial_{x}^{2} + \epsilon)\phi_{+},
\label{eq8}
\end{equation}
with $a = \bar{\beta}\beta$, $b = \epsilon\bar{\beta} + \bar{\epsilon}\beta$, $c = \epsilon\bar{\epsilon} - \chi^{2}k_{y}^{2}$, $\epsilon = E - \alpha - \gamma k_{y}^{2}$, and $\bar{\epsilon}= E - \bar{\alpha} - \bar{\gamma} k_{y}^{2}$. The general solutions of Eq.~(\ref{eq7}) have the form
\begin{equation}
\phi_{+} = \delta_{1}e^{zx} + \delta_{2}e^{-zx} + \bar{\delta}_{1}e^{\bar{z}x} + \bar{\delta}_{2}e^{-\bar{z}x},
\label{eq9}
\end{equation}
with the coefficients of the exponentials defined by
\begin{subequations}
\begin{align}
& z = \sqrt{\sqrt{\left(\frac{b}{2a}\right)^{2} - \frac{c}{a}} - \left(\frac{b}{2a}\right)} ,\label{eq10a} \\ 
& \bar{z} = \sqrt{-\sqrt{\left(\frac{b}{2a}\right)^{2} - \frac{c}{a}} - \left(\frac{b}{2a}\right)}. \label{eq10b}
\end{align}
\end{subequations}
Equation~(\ref{eq9}) has four constants ($\delta$'s) to be determined according to the boundary conditions. Replacing Eq.~(\ref{eq9}) into the relation between the first and second component of the wave function (Eq.~(\ref{eq8})), we find that
\begin{equation}
\phi_{-} = \Delta_{1}e^{zx} + \Delta_{2}e^{-zx} + \bar{\Delta}_{1}e^{\bar{z}x} + \bar{\Delta}_{2}e^{-\bar{z}x},
\label{eq11}
\end{equation}
with
\begin{equation}
\Delta_{j} = \zeta\delta_{j} , \ \ \mbox{and}  \ \ \bar{\Delta_{j}} = \bar{\zeta}\bar{\delta}_{j},
\label{eq12}
\end{equation}
for $j = 1, 2$, being $\zeta = -(i/\chi k_{y})(\epsilon + \beta z^{2})$ and $\bar{\zeta} = -(i/\chi k_{y})(\epsilon + \beta \bar{z}^{2})$. Let us admit that the ribbon is limited along the region $0 \le x \le W$. Thereby, we can write the following relations at the boundaries
\begin{equation}
\begin{aligned}
\phi_{+}(0) = {} & \delta_{1} + \delta_{2} + \bar{\delta}_{1} + \bar{\delta}_{2}, \\
\phi_{+}(W) = {} & \delta_{1}e^{zW} + \delta_{2}e^{-zW} + \bar{\delta}_{1}e^{\bar{z}W} + \bar{\delta}_{2}e^{-\bar{z}W}, \\
\phi_{-}(0) = {} & \zeta (\delta_{1} + \delta_{2}) + \bar{\zeta}(\bar{\delta}_{1} + \bar{\delta}_{2}), \\
\phi_{-}(W) = {} & \zeta(\delta_{1}e^{zW} + \delta_{2}e^{-zW}) + \bar{\zeta}(\bar{\delta}_{1}e^{\bar{z}W} + \bar{\delta}_{2}e^{-\bar{z}W}). \\
\end{aligned}
\label{eq13}
\end{equation}
\begin{figure}[!bpht]
\centerline{\includegraphics[width = 0.9\linewidth]{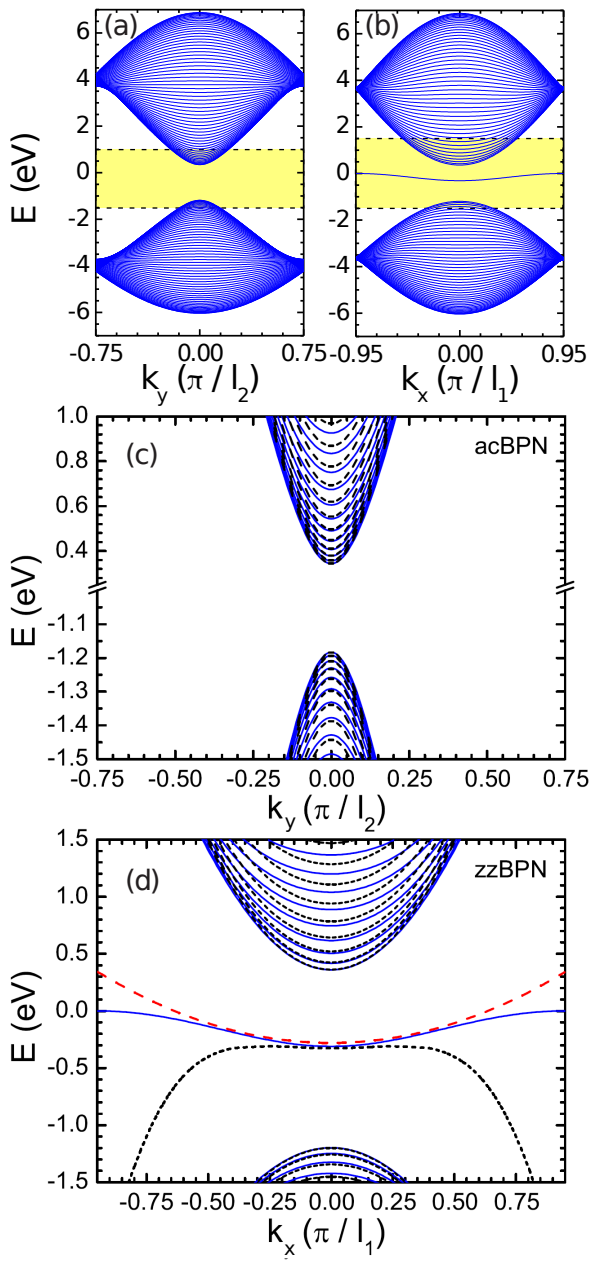}}
\caption{(Color online) Band structures for black phosphorene nanoribbons with (a, c) armchair and (b, d) zigzag edges and ribbon width $W = 101$ \AA. A comparison between the energy bands obtained by using the tight-binding model (blue solid curves) and the long-wavelength approximation (black dashed curves) is shown in (c) and (d) for acBPN and zzBPN, respectively, for the shaded yellow regions depict in (a) and (b) around $E=0$. Red dashed curve in the zigzag spectrum represents the second order approximation of the edge state reported in Ref. \onlinecite{Ezawa}.} 
\label{Fig2}
\end{figure}

Eliminating $\delta_{1}$ and $\delta_{2}$ from the above equations (\ref{eq13}), it results into
\begin{equation}
\begin{array}{c}
(\bar{\delta}_{1} + \bar{\delta}_{2})(\bar{\zeta} - \zeta) = (1 - \zeta)\mathcal{A}(0) - (1 + \zeta)\mathcal{B}(0), \\
(\bar{\delta}_{1}e^{\bar{z}W} + \bar{\delta}_{2}e^{-\bar{z}W})(\bar{\zeta} - \zeta) = (1 - \zeta)\mathcal{A}(W) - (1 + \zeta)\mathcal{B}(W),
\end{array}
\label{eq14}
\end{equation} 
where we had defined $\mathcal{A}(x) = \sum_{i = A,D}\phi_{i}(x)$ and $\mathcal{B}(x) = \sum_{i = B,C}\phi_{i}(x)$, by taking the advantage of the symmetry between the sublattices $A/D$ and $B/C$. Since, for a armchair terminated edge, we always have a pair of missing atoms of inequivalent sublattices in both edges, we can impose the boundary conditions such as
\begin{equation}\label{eq.BC.AC}
\mathcal{A}(x=0) = \mathcal{A}(x=W) = \mathcal{B}(x=0) = \mathcal{B}(x=W) = 0. 
\end{equation} 
Similar standard boundary conditions were defined for graphene nanoribbons by L. Brey \textit{et al.}\cite{Brey}. According to the authors, the appropriate boundary conditions for armchair graphene nanoribbons is for the wave function to vanish on both sublattices at the edges, i.e. the probability amplitude along the $A-B$ dimers at the edge of graphene is set to zero, since graphene is described by two triangular sublattices labelled by $A$ and $B$. In the case of an acBPN, not only the probability amplitudes of the atoms on the edges vanishes but also of it's corresponding symmetry partners. Applying the boundary conditions at Eq. (\ref{eq14}), we arrive at the following system of two algebraic equations 
\begin{equation}
\left\{
\begin{array}{l}
(\bar{\delta}_{1} + \bar{\delta}_{2})(\zeta - \bar{\zeta}) = 0, \\
(\bar{\delta}_{1}e^{\bar{z}W} + \bar{\delta}_{2}e^{-\bar{z}W})(\zeta - \bar{\zeta}) = 0.
\end{array}\right.
\label{eq15}
\end{equation}

At this point, we have to analyze two different situations: $\bar{\zeta} \neq \zeta$ and $\bar{\zeta} = \zeta$, which amounts to consider $z \neq \bar{z}$ and $z = \bar{z}$, respectively. Assuming $z = \bar{z}$ in Eq.~(\ref{eq13}), we find $(\delta_1 + \bar{\delta}_{1})\sinh(zW) = 0$. By plotting $\sinh(zW)$, no point inside the Brillouin zone is detected in which this function vanishes and, therefore, we end up with $\delta_1 + \bar{\delta}_{1} = 0$. Consequently, for this special case, we find the trivial solutions $\phi_{\pm} = 0$, which add no information to the continuum description and, therefore, are useless to our purposes. Thus, we must impose the condition $\bar{\zeta} \neq \zeta$ in order to find non-trivial solutions and to be in agreement with the tight-binding results. Applying this condition to Eq.~(\ref{eq15}), we immediately arrive at $\exp(2\bar{z}W) = 1$ and consequently at $2\bar{z}W = 2in\pi$. If $\bar{z}$ is a pure imaginary complex number, we can take $\bar{z} = ik_n$, with $k_{n} = n\pi/W$ and $n = 1, 2, 3...$, resulting in the following dispersion relation for electron and holes
\begin{equation}
E_{n} = u_{0} + \eta_{x}k_{n}^{2} + \eta_{y}k_{y}^{2} \pm \sqrt{(\delta + \gamma_{y}k_{y}^{2} + \gamma_{x}k_{n}^{2})^{2} + \chi^{2}k_{y}^{2}},
\label{eq16}
\end{equation}
where plus (minus) sign yields the conduction (valence) band. The obtained above relation corresponds to the Bulk solution of Ref.~[\onlinecite{Milton}] for $k_n \leftrightarrow k_x$ in the limit $W \rightarrow \infty$. Figure \ref{Fig2}(a) shows the band structure for armchair BPN with width $W = 101$ \AA, measured with respect the center of first Brillouin zone and obtained via tight-binding model. The spectrum highlighted in the shaded yellow region of Fig.~\ref{Fig2}(a) between the two horizontal parallel dashed lines is enlarged in Fig.~\ref{Fig2}(c) in order to compare the tight-binding results (solid curves) and the long-wavelength approximation (dashed curves). The excellent agreement between those results demonstrates that the proposed symmetry-based boundary conditions are able to describe accurately the main features of the low-energy electronic states in acBPNs. Notice from Fig.~\ref{Fig2}(a) that for higher energies, the curvature of the energy levels becomes negative. Such behaviour is not captured by our theoretical model and it may be not associated with the specific boundary conditions that we have presented here. Since our two-band model is derived from the continuum approximation reported in Ref.~[\onlinecite{Milton}], both approaches have the same limit of validity of approximately $-2.0$ eV to $1.5$ eV, when it is compared with the tight-binding results. Similar mismatch is observed for the graphene nanoribbon case, such that the analytic solution based on the boundary conditions described by Brey\cite{Brey} does not match in higher energy range \cite{Enoki, Enoki1}.
 
By analyzing the set of equations (\ref{eq15}), one can see that by rewriting them in terms of $\delta_{i}$ and $z$, with $i = 1, 2$, they lead to the condition $\exp(2zW) = 1$, resulting in a pure imaginary $z$, as obtained in the first case. However, the real part of $z$ never vanishes, consequently yielding $\exp(2zW) \neq 1$. Thus, one has to assume $\delta_{1} = \delta_{2} = 0$ in order to eliminate that choice and obtain reasonable solutions. This is never the case of the $\bar{z}$ coefficient, since its real part always vanishes for values of $E$ and $k_{y}$ inside the bulk region.

\begin{figure}[t]
\centerline{\includegraphics[width = \linewidth]{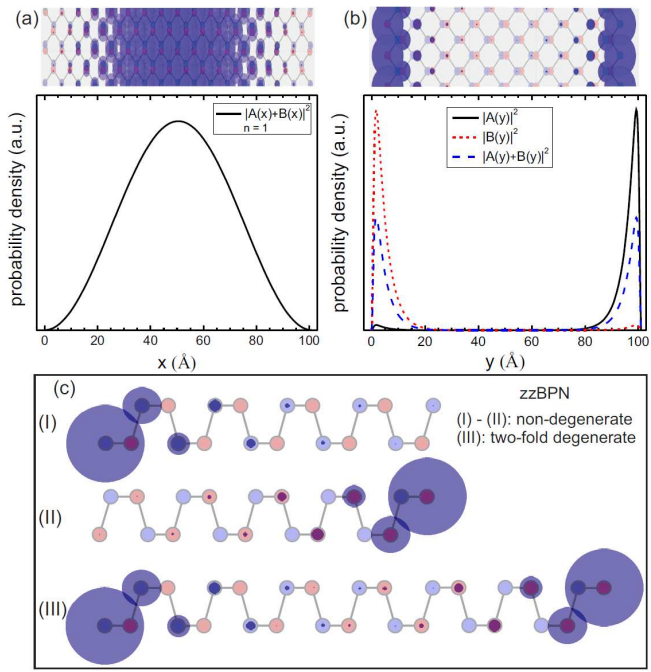}}
\caption{(Color online) Probability density for (a) acBPN and (b) zzBPN, with energies $E \approx 0.39$ eV and $E \approx -0.3$ eV, respectively, obtained from the tight-binding model (top) and the continuum approximation (bottom). Those energies corresponds to the bottom of the edge state for the zigzag case, and the bottom of the first mode ($n = 1$) state for an armchair ribbon considering $k_{y} = 0$. The blue disks in the tight-binding results denote the probability weight on the atoms. Sublattices $A$ and $D$ ($B$ and $C$) are represented by the light red (light blue) atoms. (c) Schematic cross-section view along the real space localization of the probability amplitudes for the three different combinations of zigzag terminations (I)-(III). Panels (I) and (II) show the probability density of two non-degenerate edge states being the electron localized at one edge, whereas panel (III) represents a two-fold degenerate situation.} 
\label{Fig3}
\end{figure}

The wave functions for an acBPN, corresponding to the bulk states, are linear combinations of all possible modes 
\begin{equation}
\psi = \sum_{n}A_{n}\left(
\begin{array}{c}
1 \\
\bar{\zeta}_{n}
\end{array}\right)e^{ik_{y}y}\sin\left(\frac{n\pi}{W}x\right),
\label{eq17}
\end{equation}
with $\bar{\zeta}_{n} = -(i/\chi k_{y})(\epsilon_{n} - \beta k_{n}^{2})$ and $A_{n}$ being a normalization constant. The index $n$ indicates the number of nodes of the confined wave function. In Fig.~\ref{Fig3}(a), we plot the wave function corresponding to the first energy mode $n = 1$ for wave vectors near $k_y = 0$, i.e. the electronic state associated with the conduction band minimum. A comparison between the probability amplitudes obtained from the tight-binding model (top panel) and the continuum approximation (bottom panel) is shown. The probability weight is proportional to the blue disk radius in the tight-binding result (top panel) and atoms with the same color represent sites of equivalent sublattices, being the $B/C$ ($A/D$) sites represented by light red (light blue) atoms. For visualization purposes, we have considered an acBPN with an arbitrary  width in the tight-binding plotting instead of $W=101$ \AA\ for the analytical one, so that it comports a smaller number of atomic sites along its traversal direction. From both results, we can notice that the wave function is localized at the middle of the armchair nanoribbon, i.e. this state is confined in the bulk region of the ribbon. Analogous behaviour is observed to the armchair graphene nanoribbons as reported in the literature\cite{Brey, Enoki, Enoki1}. Therefore, we can verify that our proposed boundary condition for armchair BPNs describes properly the band structure for low-energy regime with a very good agreement with the tight-binding results. Moreover, one can easily verify that the probability density at the atomic sites on the edges in the tight-binding result are in accord with the proposed boundary conditions given by Eq.~(\ref{eq.BC.AC}).  

\section{Zigzag phosphorene nanoribbons}\label{Sec.ZZ}

The geometry of a zigzag phosphorene nanoribbon (zzBPN) is illustrated at the bottom part of Fig.~\ref{Fig1}(b), where $l_{1} = 3.31$ \AA\ is the lattice parameter of the unit cell. For that orientation the edge runs along the $x$-axis and the finite width $W$ of the nanoribbon produces confinement of the electronic states along the $y$-direction. As in graphene nanoribbons with zigzag edges\cite{Brey}, zzBPNs have just one phosphorus atom in each edge. The phosphorus atoms are arranged in such a way that two different configurations for the edges are possible, depending on the ribbon width: both edges composed of atoms of equivalent sublattice, or each edge formed by atoms of inequivalent sublattices. The former configuration is not considered in our calculations, since it will generate a zzBPN with one imperfect zigzag edge (known as cliff edge\cite{Carvalho} or beard edge \cite{Ezawa}) that is subject to reconstruction. Previous theoretical \cite{Barone} and experimental\cite{Ritter} works have demonstrated interesting features coming from the relaxation and passivation of edge structures in graphene nanoribbons, which provide additional ways of modifying their electronic structure. Analogously, it has been recent reported via first-principles calculations of BPNs passivated by H atoms \cite{Tran1} and functionalized by different edge groups \cite{Peng}, such as H, F, Cl, OH, O, S, and Se, to stabilize the structures removing the edge dangling bonds. For instance, a considered case of zzBPN discussed in the present paper is sketched in Fig.~\ref{Fig1}(b), where it contains only atoms from sublattice $A$ at the bottom edge, and only atoms of sites $C$ are present at the opposite edge.   

According to Fig.~\ref{Fig1}, the zigzag nanoribbons have translational invariance along the $x$-direction that guarantees that the wave functions can be written in the form $\phi \rightarrow \phi e^{ik_x x}$. In this case, Eqs.~(\ref{eq3}) and (\ref{eq4}) become
\begin{subequations}
\begin{align}
(\gamma\partial_{y}^{2} + \epsilon)\phi_{+} &= \chi\partial_{y}\phi_{-}, \label{eq18}\\
(\bar{\gamma}\partial_{y}^{2} + \bar{\epsilon})\phi_{-} &= - \chi\partial_{y}\phi_{+}, \label{eq19}
\end{align}
\end{subequations}
where $\epsilon = E - \alpha - \beta k_{x}^{2}$ and $\bar{\epsilon} = E - \bar{\alpha} - \bar{\beta} k_{x}^{2}$. Decoupling the above system, we obtain for the component $\phi_{+}$
\begin{equation}
(a\partial_{y}^{4} + b\partial_{y}^{2} + c)\phi_{+} = 0,
\label{eq20}
\end{equation}
with $a = \bar{\gamma}\gamma$, $b = \epsilon\bar{\gamma} + \bar{\epsilon}\gamma + \chi^{2}$ and $c = \bar{\epsilon}\epsilon$. The general solutions can be written as 
\begin{subequations}
\begin{align}
\phi_{+} &= \delta_{1}e^{zy} + \delta_{2}e^{-zy} + \bar{\delta}_{1}e^{\bar{z}y} + \bar{\delta}_{2}e^{-\bar{z}y}, \label{eq22}\\
\phi_{-} &= \Delta_{1}e^{zy} + \Delta_{2}e^{-zy} + \bar{\Delta}_1e^{\bar{z}y} + \bar{\Delta}_{2}e^{-\bar{z}y}, \label{eq23}
\end{align}
\end{subequations}
with $z$ and $\bar{z}$ given by Eqs.~(\ref{eq10a}) and (\ref{eq10b}), respectively. The relation between the coefficients present in $\phi_+$ and $\phi_-$ are given by
\begin{equation}
\Delta_{1} = \zeta \delta_{1}, \quad \Delta_{2} = -\zeta \delta_{2}, \quad \bar{\Delta}_{1} = \bar{\zeta} \bar{\delta}_{1}, \quad \bar{\Delta}_{2} = -\bar{\zeta} \bar{\delta}_{2},
\label{eq24}
\end{equation}
where
\begin{equation}
\zeta = \frac{1}{\chi z}(\epsilon + \gamma z^2), \qquad \bar{\zeta} = \frac{1}{\chi \bar{z}}(\epsilon + \gamma \bar{z}^2).
\label{eq25}
\end{equation}

By evaluating the functions $\phi_{+}$ and $\phi_{-}$ in both edges, i.e. at $y = 0$ and $y = W$, for a ribbon with width $W$, the resulting set of equations become
\begin{equation}
\left[
\begin{array}{cccc}
1 & 1 & 1 & 1 \\
\zeta & - \zeta & \bar{\zeta} & -\bar{\zeta} \\
e^{zW} & e^{-zW} & e^{\bar{z}W} & e^{-\bar{z}W}\\
\zeta e^{zW} & -\zeta e^{-zW} & \bar{\zeta} e^{\bar{z}W} & - \bar{\zeta}e^{-\bar{z}W}
\end{array}
\right]
\left[
\begin{array}{c}
\delta_{1} \\
\delta_{2} \\
\bar{\delta}_{1} \\
\bar{\delta}_{2} 
\end{array}
\right] = 
\left[
\begin{array}{c}
\phi_{+}(0) \\
\phi_{-}(0) \\
\phi_{+}(W) \\
\phi_{-}(W) 
\end{array}
\right].
\label{eq26}
\end{equation}

Taking into account the symmetry between the equivalent sublattices in the different sublayers, i.e. $A/D$ and $B/C$ sites, one can rewrite $\phi_{+}$ and $\phi_{-}$ as $\phi_{+}(y) = \mathcal{A}(y) + \mathcal{B}(y)$ and $\phi_{-}(y) = \mathcal{A}(y) - \mathcal{B}(y)$, being $\mathcal{A}(y) = \sum_{i = A,D}\phi_{i}(y)$ and $\mathcal{B}(y) = \sum_{i = B,C}\phi_{i}(y)$, as considered in the previous section for the acBPN. In addition, one can easily verify that the different configurations for the edge terminations imply in different boundary conditions, which should reproduce the tight-binding results. As already mentioned, we consider the configuration shown in bottom part of Fig.~\ref{Fig1}(b), where the atom in the top (bottom) edge is from the $C$ ($A$) sublattice. In graphene nanoribbon with zigzag edges, the correct boundary condition is for the wave function to vanish on a single sublattice at each edge \cite{CastroNetoReview, Brey, Enoki, Enoki1}. Here, it seems that for zzBPN the appropriate boundary condition based on the symmetry between $A/D$ and $B/C$ atoms is for the wave function of the coupled sublattices whose atoms are not present at the edges to vanish. For instance, the boundary conditions for the case shown in Fig.~\ref{Fig1}(b) are
\begin{equation}\label{eq.BC.ZZ}
\mathcal{A}(y=0) = \mathcal{B}(y=W) = 0.
\end{equation}
Such boundary conditions allow the existence of surface states strongly localized near the edges, which are non-vanishing only on the coupled sublattices present in the edges. As a consequence of Eq.~(\ref{eq.BC.ZZ}), the matrix column on the right side of Eq.~(\ref{eq26}) is now written as $[1 \ \ -1 \ \ 1 \ \ 1]^{T}$, where we have assumed, by symmetry, that $\mathcal{A}(y=W) = \mathcal{B}(y=0)$ and that their value equal 1, for simplicity. Hence, using such conditions, the solutions of the non-homogeneous linear system of equations (\ref{eq26}) are
\begin{subequations}
\begin{align}
\delta_{1} &= \frac{1}{\Omega}[\bar{\zeta}(e^{\bar{z}W} - 1) - (e^{\bar{z}W} + 1)],\label{eq27}\\
\bar{\delta}_{1} &= \frac{1}{\Omega}[\zeta(1 - e^{zW}) + (1 + e^{zW})], \label{eq28}\\
\delta_{2} &= e^{zW}\delta_{1}, \ \mbox{~~~~~~~~~} \ \bar{\delta}_{2} = e^{\bar{z}W}\bar{\delta}_{1},\label{eq29}
\end{align}
\end{subequations}
with
\begin{equation}
\Omega = \bar{\zeta}(e^{zW} + 1)(e^{\bar{z}W} - 1) - \zeta(e^{\bar{z}W} + 1)(e^{zW} - 1).
\label{eq30}
\end{equation}
Since the boundary conditions (Eq.~(\ref{eq.BC.ZZ})) are applied into the functions $\mathcal{A}(y)$ and $\mathcal{B}(y)$, it is more convenient to rewrite them by using the definitions (\ref{eq22}) and (\ref{eq23}), such as
\begin{subequations}
\begin{align}
2\mathcal{A}(y) = & ~\delta_{1}(1 + \zeta)e^{zy} + \delta_{2}(1 - \zeta)e^{-zy} + \nonumber \\ & ~\bar{\delta}_{1}(1 + \bar{\zeta})e^{\bar{z}y} + \bar{\delta}_{2}(1 - \bar{\zeta})e^{-\bar{z}y},\label{eq31}\\
2\mathcal{B}(y) = & ~\delta_{1}(1 - \zeta)e^{zy} + \delta_{2}(1 + \zeta)e^{-zy} + \nonumber \\ & ~\bar{\delta}_{1}(1 - \bar{\zeta})e^{\bar{z}y} + \bar{\delta}_{2}(1 + \bar{\zeta})e^{-\bar{z}y}\label{eq32}.
\end{align}
\end{subequations}
Applying Eq.~(\ref{eq.BC.ZZ}) into the above equations, we obtain after some algebraic manipulations 
\begin{subequations}
\begin{align}
0 = & ~(\delta_{1} + e^{-zW}\delta_{2})((1 + \zeta) + (1 - \zeta)e^{zW}) + \nonumber \\ & ~(\bar{\delta}_{1} + e^{-\bar{z}W}\bar{\delta}_{2})((1 + \bar{\zeta}) + (1 - \bar{\zeta})e^{\bar{z}W}),\label{eq35}\\
0 = & ~(\delta_{1} - e^{-zW}\delta_{2})((1 + \zeta) - (1 - \zeta)e^{zW}) + \nonumber \\ & ~(\bar{\delta}_{1} - e^{-\bar{z}W}\bar{\delta}_{2})((1 + \bar{\zeta}) - (1 - \bar{\zeta})e^{\bar{z}W}).\label{eq36}
\end{align}
\end{subequations}
Eliminating the coefficients $\delta_{1}$ and $\bar{\delta}_{1}$ by inserting  Eq.~(\ref{eq29}) into Eqs.~(\ref{eq35}) and (\ref{eq36}), it allows us to rewrite the set of above equations as
\begin{align}
0 = &\delta_{2}[(1 + \zeta)e^{-zW} + (1 - \zeta)] +  \bar{\delta}_{2}[(1 + \bar{\zeta})e^{-\bar{z}W} + (1 - \bar{\zeta})].
\label{eq37}
\end{align}
We have now obtained a single transcendental equation for the full energy spectrum of the zzBPNs, in contrast to the case of acBPNs, in which we find just an explicit expression for the bulk energy states, as demonstrated in Eq.~(\ref{eq16}). In order to solve numerically Eq.~(\ref{eq37}), we impose the following constraint: $\bar{\delta}_{2} + \delta_{2} = 0$. This ansatz can be justified by the fact that, since $\delta_{2}$ and $\bar{\delta}_{2}$ are functions of $E$ and $k_{x}$, it will generate a second equation which is consistent with the resulting transcendental equation, Eq.~(\ref{eq37}), after the substitution $\delta_{2} = - \bar{\delta}_{2}$. To be more specific, constraints of the general form $\bar{\delta}_{2} + \mu \delta_{2} = 0$, where $\mu$ is a constant, are the only constraints that create a second equation which is consistent with the resulting Eq.~(\ref{eq37}). The particular choice $\mu = 1$ is taken to fit the bottom of the quasi-flat band (at the $\Gamma$ point) with the one obtained via tight-binding model. Explicitly speaking, we have in summary that: (i) the $\bar{\delta}_2+\delta_2 =0$ assumption was not derived from a previous condition, and indeed it was imposed to reproduce the electronic states of the zzBPN obatined within the tight-binding model; and (ii) it was the simpler condition we found in order to fit the tight-binding results without mathematical inconsistencies. Similarly to the acBPN case, we can write down the bulk dispersion relation for zzBPN as
\begin{equation}
E_{n} = u_{0} + \eta_{x}k_{x}^{2} + \eta_{y}k_{n}^{2} \pm \sqrt{(\delta + \gamma_{y}k_{n}^{2} + \gamma_{x}k_{x}^{2})^{2} + \chi^{2}k_{n}^{2}},
\label{eq38}
\end{equation}
where $k_{n} = n\pi/W$ and $n=1,2,3\ldots$.

Figure \ref{Fig2}(b) shows the energy dispersion relation for a zzBPN obtained via tight-binding model for a ribbon with $W=101$ \AA. The highlighted region (yellow rectangle delimited by dashed black lines) is enlarged and depicted in Fig.~\ref{Fig2}(d) for a better comparison between the results obtained via the tight-binding model (blue solid curves) and the continuum approximation (black dashed curves). Our analytical results for the bulk states shows a good agreement with the tight-binding ones that improves for ribbons with greater widths. In addition, unlike the armchair case, the tight-binding results for zzBPNs shown in Figs.~\ref{Fig2}(b) and \ref{Fig2}(d) exhibit an additional state between the conduction and valence bands, which corresponds to states localized on the edges of the ribbon. Such additional feature of zzBPN emerges naturally in the analytical description as a consequence of the boundary conditions, as shown in Fig.~\ref{Fig2}(d), and diverges from the tight-binding one for $k_{x}$ values away from the $\Gamma$ point. The limited range of $k_{x}$ values in which both models match corresponds to those values inside the region where the analytical edge state obtained from the continuum approximation fits the tight-binding result in Fig.~\ref{Fig2}(d). Based on theses results, we estimate the wavelength limit of validity of the analytical edge states as $\lambda \gtrsim 3.31$ nm. For comparison purposes, a moderate doping of $\approx 2 \times 10^{12}$ cm$^{-2}$ would provide a fermi wavelength of $\approx 18$ nm in a two dimensional semiconductor such as phosphorene \cite{PNAS}.

In order to analyze in more details the analytical result for the edge states, we compare this result with the studies reported recently by Esawa in Ref.~[\onlinecite{Ezawa}] for the quasi-flat band in the anisotropic honeycomb-lattice model. Based on the previous knowledge of the existence of such edge state, Ezawa has explored the origin of the quasi-flat band, investigating the changes of the band structures of BPNs by modifying the hopping parameters, and estimated perturbatively an approximate analytical expression for the energy spectrum of the quasi-flat band. This expression can be written as $E(k_{x}) = -(4t_1t_4/t_2)\left[1 + \cos(k_{x}l_{1})\right]$, where $4t_1t_4/t_2 \approx -0.14$ eV. By expanding this relation up to the second order owing to a direct comparison with our analytical result for the edge state, we have that $E(k_{x}) \approx - 0.28\left[1 - (k_{x}l_{1}/2)^{2}\right]$. Shown by the red dashed curve in Fig.~\ref{Fig2}(d), the result of Ref.~[\onlinecite{Ezawa}] has the advantage to fit the tight-binding result for a larger wave vector range, but on the other hand, it was obtained by the previous knowledge of the existence of the flat band, whereas here it naturally arise from the analytical calculation. These results can be improved by extending the continuum approximate Hamiltonian with terms up to third or even higher order to $k$ vector, instead of the one given by Eq.~(\ref{eq1}). However, the new continuum Hamiltonian would lead us to more complicated equations in which would be unnecessary, since the results have demonstrated that the present model is sufficient to describe the main features of phosphorene nanoribbons at the wavelength limit reached in experimental basis. 

Another important comparison that needed to be done is concerning the recent obtained first-principles calculation results for BP nanoribbons with different types of edges, such as zigzag, armchair, and cliff edges by considering or not the possibility of reconstructed edges, as well as by saturating the edges with hydrogen passivation. In this perspective, Carvalho \textit{et al.}\cite{Carvalho} found out that the nature of edge-induced gap states on BPNs depends not only on the crystal structure of how BP sheet is cut, but also on the way it terminates, in a similar way to our obtained results here and, on the other hand, that their band structures can be exhibit metallic or semiconductor behavior if the atoms on the BPN edges undergo a reconstruction or distortion. Their findings had shown that all stable BP nanoribbons with unsaturated edges (i.e. zigzag, cliff and even armchair edges) have edge-induced gap states that can be removed by hydrogen passivation. It is due to the fact that the BPNs with armchair and cliff edges in their calculations are allowed to be reconstructed with the lengths of phosphorus bounds on the edges being different of the non-deformed part of BPN. This is in contrast to our armchair spectrum (Fig. \ref{Fig2}(c)), where the middle gap states are absent. However, similarly to Fig. \ref{Fig2}(d), they also found that these zigzag nanoribbons have a two-fold degenerate edge-related states for larger ribbon width and that the dispersion very close to $\Gamma$ point of these states are approximately concave-up parabolas.

Figure \ref{Fig3}(b) shows the probability density for the edge states of a zzBPN obtained using tight-binding model (top panel in Fig.~\ref{Fig3}(b)) and continuum approximation (bottom panel in Fig.~\ref{Fig3}(b)). These results confirm that these are nodeless confined states localized at the edges of the nanoribbon. For the tight-binding result, we have considered a ribbon with arbitrary smaller width in order to have a clearer representative BPN, instead of the ribbon width $W=101$ \AA\ taken for the analytical case. Similarly to the acBPN results shown in Fig.~\ref{Fig3}(a), the size of the blue disk radius is related to the probability amplitude of the squared wave function and the equivalent sublattices are represented with the same color for the tight-binding atomic structures. From the continuum result (bottom panel in Fig.~\ref{Fig3}(b)), one can note that $|\mathcal{A}(y)|^2$ is localized near the edge on $y = W$ (black solid curve), whereas $|\mathcal{B}(y)|^2$ is confined around $y = 0$ (red dashed curve), as expected due to the boundary condition Eq.~(\ref{eq.BC.ZZ}). This way, the total probability amplitude $|\mathcal{A}(y)+\mathcal{B}(y)|^2$ is distributed along both zigzag edges (blue long dashed curve). On the other hand, one can also notice that the total probability density obtained from the tight-binding model, as shown in top panel of Fig.~\ref{Fig3}(b) for a representative ribbon width, exhibits well-localized states on both zigzag edges. Furthermore, we can verify, by taking a more careful look in the probability density per site, that close to the $y = W$ edge the amplitudes are non-zero only for atomic sites of sublattices $A$ and $D$ (represented by the blue atoms), whereas, close to the opposite edge $y=0$, the amplitudes are centered exclusively on sites of sublattices $B$ and $C$ (illustrated by the red atoms). This tight-binding result is in agreement with the long-wavelength description, and thus supports the symmetry-based boundary conditions proposed here. 

In addition, similar boundary conditions (equivalently, see the boundary conditions given by Eq. (\ref{eq.BC.ZZ}) for zigzag-zigzag BPNs) from the perspective of the continuum model can be set by zigzag-beard and beard-zigzag edge terminations. When the nanoribbon lattice is cut in a way that one of the boundaries has a beard termination, the charge carriers avoid the beard edge and are mostly confined along the zigzag edge that is located on the opposite boundary. Therefore, the appropriate boundary condition for this system based on the symmetry between $A/D$ and $B/C$ is for both wave functions of the coupled sublattices $B/C$ and $A/D$ to vanish along the cliff edge and at the coupled sublattices whose atoms are not present at the edges. Let us consider the configuration shown in the bottom part of Fig. \ref{Fig1}(b) where the atom in the top (bottom) edge is from the $C$ ($A$) sublattice, but instead of both edges being zigzag edges, we assume that one of them is beard. For instance, if the bottom (top) edge is beard, then the appropriate boundary condition for this system is for $\phi_{A,B,C,D}(y=W) = \phi_{A,D}(y=0) = 0$ [$\phi_{A,B,C,D}(y=0) = \phi_{B,C}(y=W) = 0$], that implies $\mathcal{A}(y=W) = \mathcal{B}(y=W) = \mathcal{A}(y=0) = 0$ [$\mathcal{A}(y=0) = \mathcal{B}(y=0) = \mathcal{B}(y=W) = 0$]. By using the mentioned boundary conditions for beard-zigzag (zigzag-beard) edges and following the mathematical procedure developed by Eqs. (\ref{eq.BC.ZZ})-(\ref{eq37}), we can find a equivalent transcendental equation for the bulk and edge energy levels in case of the boundaries with beard edge. According to the obtained boundary conditions, one finds that the probability densities for the beard BPNs in the continuum approximation are similar to the already plotted at the bottom panel in Fig. \ref{Fig3}(b), where $|\mathcal{B}(y)|^2$ ($|\mathcal{A}(y)|^2$) is now localized near the only zigzag edge on $y = 0$ ($y=W$) for beard-zigzag (zigzag-beard) BPNs, such that the total amplitude contribution is due only the state on the zigzag edge, yielding this way a single non-degenerate edge state in the energy spectrum.

Figure~\ref{Fig3}(c) displays the probability amplitudes over schematic cross-section views of the atomic structures for the three non-equivalent combinations of zigzag terminations in BPNs, labelled by (I), (II) and (III). (I) and (II) represent two different zigzag-beard terminations for zzBPNs with edge atoms composed by the same sublattice coupled group (light blue - light blue or light red - light red), whereas panel (III) shows a zigzag-zigzag phosphorene nanoribbon with the edge atoms formed by different sublattice group type (light blue - light red). These plots show in which of the possible zigzag terminations the electronic structure exhibit degenerate or non-degenerate edge states. The band structure of BPNs was investigated in Ref.~[\onlinecite{Ezawa}] for three types of terminations, whose edges are both zigzag, zigzag and beard, and both beard. It showed that the presence of the quasi-flat edge modes isolated from the bulk modes are doubly degenerate for a zigzag-zigzag nanoribbon, and non-degenerate for a zigzag-beard nanoribbon, while they are absent in a beard-beard nanoribbon. A similar behaviour is observed in the tight-binding results, where panels (I) and (II) show the probability density of two non-degenerate edge states being the electron localized at one edge, whereas panel (III) represents a two-fold degenerate situation.

A new type of edges for BPNs were recently proposed\cite{Marko} by taking the advantage of $C_2$ symmetry of the puckered BP structure, that is due to its anisotropic lattice. These new edges were called skewed edges and the respective nanoribbons as skewed-zigzag (s-zz) and skewed-armchair (s-ac) nanoribbons, which may be obtained by cutting the BP sheet in such a way that the zigzag (armchair) direction intersects the puckered ridges from a different angle than $0$ ($90$) degree. These skewed nanoribbons had shown an unexpected duality behavior as compared to the nanoribbons with normal edges, whose nature has a topological origin, finding for instance that the s-zz (s-ac) nanoribbons are semiconducting (metallic), while normal zz (ac) nanoribbons exhibit opposite features. This way, equivalent boundary conditions for the skewed nanoribbons as the ones propose in Eq. (\ref{eq.BC.ZZ}) can be addressed, since: (i) the coupled sublattice $A/D$ and $B/C$ symmetries still hold for skewed nanoribbons; (ii)  just certain atoms are present at the boundaries: the atoms on the skewed zigzag (skewed armchair) edges belong to different (the same) coupled $A/D$ and $B/C$ sublattice symmetry, such that the boundary conditions for these BPNs are similar to the ones for BPNs with normal armchair (zigzag) edges. Since we have a pair of missing atoms of inequivalent (equivalent) sublattices in both edges for skewed zigzag (skewed armchair) BPNs, we can impose the boundary conditions, such as $\mathcal{A}(y'=0) = \mathcal{A}(y'=W) = \mathcal{B}(y'=0) = \mathcal{B}(y'=W) = 0$ [$\mathcal{A}(x'=0) = \mathcal{B}(x'=W) = 0$] for skewed zigzag (skewed armchair) edges, where $x'$ and $y'$ are the new system coordinates being rotated from the normal BPN system. Consequently, no edge states are present in the skewed zigzag nanoribbons, whereas two quasiflat bands appear in the middle of the band gap for the band structure of the skewed armchair nanoribbons (see Ref. \onlinecite{Marko} for more details). This is highly contrasted to normal BP nanoribbons, and thus, it is possible to verify that the topological origin of the edge-localized states in skewed armchair and normal zigzag BPNs can be captured within our continuum model by using the correct boundary condition described along this paper.

\section{Scaling laws of band gaps for phosphorene nanoribbons}\label{Sec.GAP}

A relevant electronic property for purposes of optical applications and quantum confinement effect is related to the scaling behaviour of the band gap with ribbon width, as previously investigated in the literature for 2D materials, for instance: graphene nanoribbons\cite{CastroNetoReview, Misha1, Brey, Enoki, Enoki1, Louie, Louie1}, boron nitride nanoribbons\cite{Louie2, Guo}, silicene nanoribbons\cite{silicene1, silicene2, silicene3, silicene4} and phosphorene nanoribbons\cite{Tran1, Ezawa, Sisakht}. Recent studies via first principles calculations have indicated that the band gaps of BPNs possess different scaling laws depending on the edge type and thus suggesting its usage as a convenient tool for identifying acBPNs and zzBPNs samples with similar geometric widths, since the previous results have shown that the band gap is larger in zzBPNs than in acBPNs for the same ribbon width. Here, we analytically demonstrated for the first time the dependence of the band gap with respect to the ribbon width for BPNs with zigzag $E^{ZZ}_{g}$ and armchair $E^{AC}_{g}$ edges. Based on our theoretical model, we calculate the energy gap derived from the energy spectrum (Eqs.~(\ref{eq16}) and (\ref{eq38}) for acBPNs and zzBPNs, respectively) by selecting $k_{y(x)} = 0$ (i.e. the $\Gamma$ point) and by taking the difference between the $n = 1$ levels of the conductance and valence bands, given by
\begin{figure}[b]
\centerline{\includegraphics[width = 0.9\linewidth]{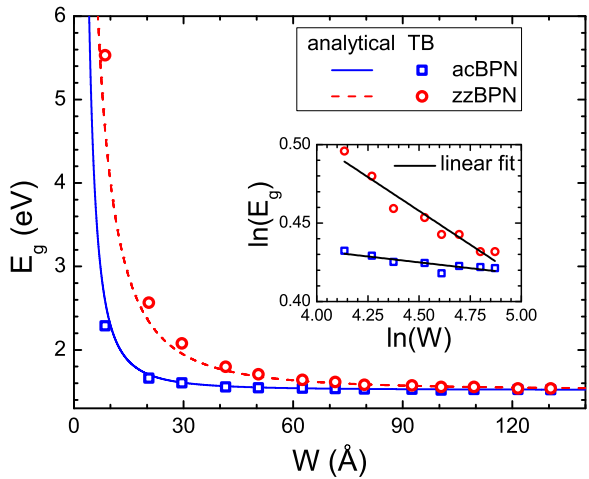}}
\caption{(Color online) Comparison between the band gap energies obtained by using the tight-binding model (symbols) and derived from the continuum approximation (curves) as a function of the nanoribbon width $W$ for the armchair (blue square-like symbols and blue solid curve) and zigzag (red circles and red dashed curve) BPN cases. The inset shows the gap energies in a logarithmic scale, where the solid black curves represent linear fits of the tight-binding results.} 
\label{Fig4}
\end{figure}
\begin{subequations}
\begin{align}
E_{g}^{AC}(W) &= 2\left(\delta + \frac{\gamma_{x}\pi^{2}}{W^{2}}\right),\label{eq39}\\
E_{g}^{ZZ}(W) &= 2\sqrt{\left(\delta + \frac{\gamma_{y}\pi^{2}}{W^{2}}\right)^{2} + \left(\frac{\chi \pi}{W}\right)^{2}},\label{eq40}
\end{align}
\end{subequations}
for armchair and zigzag phosphorene nanoribbons, respectively. In Fig.~\ref{Fig4}, we plot these band gap energies by both analytical estimate (curves) and tight-binding model (symbols) for zzBPNs and acBPNs with different ribbon widths, in order to visualize a direct comparison between the two approaches and the two different edge type. One can note from these expressions (\ref{eq39}) and (\ref{eq40}), and Fig.~\ref{Fig4} that: (i) $E_{g}$ clearly has a different behaviour for the two different types of edges discussed; (ii) for large values of $W$, the energy gap of zzBPNs scales as $\approx 1/W$, whereas the armchair ones follows a $\approx 1/W^{2}$ relation. This behaviour is observed in a clearer way in the inset of Fig.~\ref{Fig4}, where we plot the logarithm of the energy gap as a function of the logarithm of the ribbon width. Two straight lines with different inclinations fit the tight-binding results for large widths, as we would expect, since $\ln(E_g^{AC})\propto -2\ln(W)$ and $\ln(E_g^{ZZ})\propto -\ln(W)$; (iii) a consequence of the larger band gap for zzBPNs, as compared with the acBPNs for the same ribbon width, is a quantum confinement effect in zzBPNs that is more pronounced than in acBPNs; (iv) the analytical estimate is in agreement with the tight-binding calculations, specially for large ribbon widths where the continuum approximation describes with a higher accuracy the charge carriers in BPNs. The scaling laws observed in the results can be understood as a relativistic-like and nonrelativistic-like character for electrons and holes in acBPNs and zzBPNs, respectively. It is the energy-momentum dispersion relation, evidenced by which is proportional to the square of the momentum for acBPNs (see Eq.~(\ref{eq16})) whereas it is proportional to the momentum for zzBPNs (see Eq.~(\ref{eq38})). \cite{Tran1, Sisakht}  

\section{Conclusions}\label{Sec.Conclusion}

In summary, we have studied the electronic properties of single layer black phosphorene nanoribbons by using a recent proposed two-band Hamiltonian in the long-wavelength limit. Within the continuum approximation, we derived the appropriate boundary conditions to describe zigzag and armchair edges by taking advantage of the sublattice symmetry that couples the sublattices belonging to different sublayers, i.e. $A/D$ and $B/C$, being $A$ and $B$ ($C$ and $D$) in the same (top) sublayer. The analytical results were compared with those obtained via a five hopping-parameters tight-binding model, showing that both approaches may provide similar results in a given wavelength range. We estimated the wavelength range of validity of the analytical edge states and showed to be in the same experimental order as obtained in semiconductor measurements, while for the bulk states this limit is much higher, improving for larger ribbon width. We have also shown that the surface states of zigzag BPNs emerge naturally in this analytical model as a consequence of the proposed boundary conditions and that they can be confined along the both or just one zigzag edges, whereas armchair BPNs have no surface states. For the first time, we analytically computed the energy band gaps of BPNs for both edges, i.e. zigzag and armchair edges. We demonstrated that the energy gap scales differently depending on the edge type, obeying a $\approx 1/W$ ($\approx 1/W^2$) tendency for zigzag (armchair) BPNs. The analytical analysis developed along the present paper shows the possibility to use the continuum approximation to obtain accurate and relevant quantitative results especially for large phosphorene systems, which would be computationally expensive if performed using the tight-binding model.

Recently, ten-hopping tight-binding results for phosphorene nanoribbons have been reported in literature by Ref. \onlinecite{Marko}, and shown similar features for both bulk and edge states, as the ones observed in the five-hopping tight-binding assumed in this paper, such as concavity of bulk and edge energy levels, same number of bulk states and two-fold degeneracy of edge states. The only direct consequence on the continuum approach to describe carrier charge in BPNs by taking in account the assumption of more hopping energies is the parameter changes, since the parameters of the long-wavelength Hamiltonian given by Eq. (\ref{PrincipalHamiltonian}) (such as $\delta$, $\gamma_x$,	 $\gamma_y$, $\eta_x$, $\eta_y$, $\chi$, $u_0$) depend on the hopping energies. Therefore, the boundary conditions for tight-binding models with more than five-hopping parameters will have the same format as the found conditions in this work, but with different Hamiltonian parameters. In this sense, our continuum approximation captures the main features of tight-binding models already reported in the literature for BPNs and consequently the obtained boundary conditions describe in appropriate way the wave functions and the energy states in BP nanoribbons.

\section*{ACKNOWLEDGMENTS} 

This work was financially supported by the Brazilian Council for Research (CNPq), under the PRONEX/FUNCAP and CAPES foundation.

\end{document}